\documentstyle[aps,prl,twocolumn,epsfig]{revtex}
\begin{document}
%---------------------------------------------------
\title{An Electroweak Model With Massive Neutrinos}

\vskip 0.8in

\author{{Ning Wu}\\
{\small Division 1, Institute of High Energy Physics, P.O.Box 918-1,
Beijing 100039, P.R.China}}
\maketitle
\vskip 0.8in

\noindent
PACS Numbers: 14.60.Pq, 13.15+g, 12.15-y, 11.15-q, 12.10-g\\
Keywords: neutrino mass, electroweak interactions,  gauge symmetry \\
\vskip 0.3in

\noindent
\begin{abstract}
{\bf  [Abstract] }
In this paper, an electroweak model with massive neutrinos is proposed.
Because of the mixing of neutrinos, the conversion of one lepton type to
another is possible, but many of those kind processes are of
extremely small possibilities. The total
lepton-number conservation still holds in this model, so neutrinoless
double beta decay is prohibited. Though right-hand neutrinos are introduced
into the model, their coupling with matter fields is extremely small and
therefore they are much harder to be detected in experiments.

\end{abstract}

\vskip 0.3in

%-------------------------------------------------------
Recently, strong evidence that neutrinos do have nonzero masses  has been
found[1-4].
But in the Standard Model, neutrinos are massless. In order to explain
neutrino oscillation, some scenarios were put forward[5,6].
But in most of the  models which contain massive neutrinos,
the weak interactions of leptons are quite different from
those of quarks. If the Nature has beautiful symmetry, quarks and leptons
should have quite similar properties in weak interactions, and
the fundamental theories on the electroweak interactions of quarks and
leptons should have the same structure.
On the other hand, it is known that, in the Standard Model,
all quarks have right-hand components but only charged
leptons have right-hand components. What cause this asymmetry? In another
words, why neutrinos have no right-hand components while all quarks have
right-hand components?
A fundamental theory should not have so much arbitrariness and the
fundamental rules for weak interactions of quarks and leptons should
be the same.
On the basis of this consideration, we will
propose a new electroweak model in this paper.
This model inherits most of the dynamical
properties of the Standard Model but contains massive Dirac neutrinos and
is consistent with experiments on weak interactions. More important,
according to this model, the fundamental rules on the electroweak
interactions
of quarks and leptons are the same,
and the fundamental theories on their electroweak interactions
have the same structure in form[7]. As an example, we only discuss leptons
in this paper.
\\

In order to do this in a more fundamental manner, let's discuss the symmetry
of the Nature first. It is known that the
underlying symmetries of the Nature take a
fundamental role in constructing the Standard Model. Two underlying
symmetries are used in the Standard Model,
they are $SU(2)_L$ and $U(1)_Y$[8-10].
But there is another underlying symmetry of the Nature which is not
fully used in
constructing the Standard Model. This symmetry corresponds to one of the
most strict conservation laws of the Nature which had been strictly tested
by
experiments: the total lepton number conservation and total baryon number
conservation. Because leptons and quarks are usually called matter fields,
we
call this conservation matter number conservation.
This is a $U(1)$ symmetry and is denoted as $U(1)_M$ at
present. So, the symmetry of this model is supposed to be
$SU(2)_L \times U(1)_Y \times U(1)_M $. We supposed that all leptons and
quarks carry matter number $1$, all anti-leptons and anti-quarks
carry matter number $-1$, and vacuum carries no matter number.
Correspondingly, we need two kinds of gauge
fields $F_{1 \mu}$ and $F_{2 \mu}$ which correspond to
$SU(2)_L$ symmetry,  two kinds of gauge fields $B_{1 \mu}$ and
$B_{2 \mu}$ which correspond to $U(1)_Y$ and  $U(1)_M$ symmetry
respectively, and a third $U(1)$ gauge field $B_{3 \mu}$ which connects
thess two $U(1)$ symmetries.
The coupling between leptons and  gauge field $B_{1 \mu}$ is
determined by supercharge $Y$ and the coupling between leptons and
gauge field $B_{2 \mu}$ is determined by
matter number  $M$. Under $U(1)_Y$ gauge
transformations, two $U(1)$ gauge fields $B_{1 \mu}$ and $B_{3 \mu}$
will have to transform accordingly, while under  $U(1)_M$ gauge
transformations, two $U(1)$ gauge fields $B_{2 \mu}$ and $B_{3 \mu}$
will have to transform.
So, there are five different kinds of gauge covariant derivatives:
$$
\begin{array}{l}
D_{1 \mu} = \partial_{\mu} - i g F_{1 \mu},  \\
D_{2 \mu} = \partial_{\mu} + i g {\rm tg} \alpha F_{2 \mu},  \\
D_{3 \mu} = \partial_{\mu} - i g_1 B_{1 \mu} \frac{Y}{2},  \\
D_{4 \mu} = \partial_{\mu} - i g_2 B_{1 \mu} \frac{M}{2},  \\
D_{5 \mu} = \partial_{\mu} + i g' {\rm tg} \alpha B_{2 \mu} ,
\end{array}
\eqno{(1)}
$$
where $\alpha$ is a dimensionless parameter and
$$
g'=\sqrt{g_1^2+g_2^2}.
\eqno{(2)}
$$
\\

Let $e^{(i)}$ represent $e,\mu$ or $\tau$,
and $\nu^{(i)}$ represent the corresponding neutrinos
$\nu _e,\nu_{\mu}$ or $\nu _{\tau}$. That is
$$
\begin{array}{l}
e^{(1)} =e,~~e^{(2)}=\mu,~~e^{(3)}=\tau, \\
nu^{(1)}=\nu _e,~,\nu^{(2)}=\nu _{\mu},~\nu^{(3)}=\nu _{\tau}.
\end{array}
\eqno{(3)}
$$
Leptons $e^{(i)}$ and $\nu^{(i)}$ form left-hand doublets
$\psi^{(i)}_L$ and right-hand singlets $e^{(i)}_R$ and $\nu^{(i)}_R$. The
definitions of $\psi_L$, $e_R$ and $\nu_R$
are the same as those in the Standard Model. The mixing of neutrinos
is accomplished  through:
$$
\left (
\begin{array}{c}
\nu^{(1)}_{\theta} \\
\nu^{(2)}_{\theta}  \\
\nu^{(3)}_{\theta}
\end{array}
\right )
= K
\left (
\begin{array}{c}
\nu^{(1)}\\
\nu^{(2)}\\
\nu^{(3)}
\end{array}
\right ),
\eqno{(4)}
$$
where K is the mixing matrix for neutrinos.
\\

The Lagrangian of the model is
$$
{\cal L} = {\cal L} _l + {\cal L} _g + {\cal L} _{v-l}
\eqno{(5)}
$$
$$
\begin{array}{l}
{\cal L }_l= -\sum_{j=1}^3 \overline{\psi}^{(j)}_L \gamma ^{\mu}
(\partial _{\mu}
 -ig F_{1 \mu} \\
+ \frac{i}{2} (g_1 B_{1 \mu}  - g_2  B_{2 \mu}
)) \psi^{(j)}_L  \\
-\sum_{j=1}^3 \overline{e}^{(j)}_R \gamma ^{\mu}
(\partial _{\mu}+ i (  g_1 B_{1 \mu}
- \frac{g_2}{2}  B_{2 \mu} )) e^{(j)}_R  \\
 -\sum_{j=1}^3 \overline{\nu}^{(j)}_{ \theta R} \gamma ^{\mu}
(\partial _{\mu} - \frac{i}{2} g_2  B_{2 \mu} )
\nu^{(j)}_{\theta R}
\end{array}
\eqno{(6)}
$$
$$
\begin{array}{l}
{\cal L}_g = -\frac{1}{4}  F^{i \mu \nu}_1 F^i_{1 \mu \nu}
- \frac{1}{4} F^{i \mu \nu}_2 F^i_{2 \mu \nu}
-\frac{1}{4}  B^{\mu \nu}_1 B_{1 \mu \nu}  \\
-\frac{1}{4}  B^{\mu \nu}_2 B_{2 \mu \nu}
-\frac{1}{4}  B^{\mu \nu}_3 B_{3 \mu \nu}   \\
 - v^{\dag}
\lbrack  {\rm cos} \theta _W
( {\rm cos} \alpha F_1^{\mu}+{\rm sin}\alpha F_2^{\mu})  \\
 - {\rm sin}\theta _W ( {\rm cos}\alpha
({\rm cos}\beta B_{1 \mu} - {\rm sin}\beta B_{2 \mu} )
+{\rm sin}\alpha B_3^{\mu} ) \rbrack  \\
\cdot
\lbrack  {\rm cos} \theta _W ( {\rm cos}
\alpha F_{1 \mu}
+{\rm sin}\alpha F_{2 \mu})  \\
-{\rm sin} \theta _W ( {\rm cos}\alpha
({\rm cos}\beta B_{1 \mu} - {\rm sin}\beta B_{2 \mu} )
+{\rm sin}\alpha B_{3 \mu} ) \rbrack
v
\end{array}
\eqno{(7)}
$$
$$
\begin{array}{l}
{\cal L} _{v-l}= -\sum_{j=1}^3
( f^{(j)} \overline{e}_R v^{\dag} \psi^{(j)} _L
+ f^{(j) \ast} \overline{\psi}^{(j)}_L v e_R)  \\
-\sum_{j,k=1}^3 ( f^{(jk)} \overline{\psi}^{(j)}_L
\overline{v} \nu^{k}_{\theta R}
+ f^{(jk) \ast} \overline{\nu}^{k}_{\theta R} \overline{v}^{\dag}
\psi^{(j)}_L ),
\end{array}
\eqno{(8)}
$$
where
$$
\begin{array}{l}
F_{1 \mu \nu}^i = \partial _{\mu} F_{1 \nu}^i - \partial _{\nu} F_{1 \mu}^i
+g \epsilon _{ijk} F_{1 \mu}^j    F_{1 \nu}^k  \\
F_{2 \mu \nu}^i = \partial _{\mu} F_{2 \nu}^i - \partial _{\nu} F_{2 \mu}^i
-g {\rm tg} \alpha \epsilon _{ijk} F_{2 \mu}^j    F_{2 \nu}^k  \\
B_{m \mu \nu}= \partial _{\mu} B_{m \nu}- \partial _{\nu} B_{m \mu}
~~ (m=1,2,3),
\end{array}
\eqno{(9)}
$$
$v$ is a two-dimensional vacuum potential, $\overline{v}$ is given by
$$
\overline{v} = i \sigma_{2} v^{\ast} =
\left (
\begin{array}{c}
v_2^{\dag} \\
- v_1^{\dag}
\end{array}
\right ),
\eqno{(10)}
$$
and $\beta$ is a dimensionless parameter which is given by:
$$
{\rm sin} \beta = g_2 / g' ~~,~~ {\rm cos} \beta = g_1 / g'.
\eqno{(11)}
$$
It could be strictly proved that the above Lagrangian has strict local
$SU(2)_L \times U(1)_Y \times U(1)_M$ gauge symmetry[7]. \\

The symmetry breaking of the model is accomplished through the phase
transition of the vacuum. We could suppose that the above Lagrangian
describes a special state of matter which exists in the
condition of extreme high temperature. This state is a special
phase of vacuum and may exist at a moment of Big Bang.
Alone with the decreasing of the temperature of the state, the phase
transition of the vacuum occurs. After phase transition, the vacuum
potential changes into:
$$
v =\left (
\begin{array}{c}
0\\
\mu / \sqrt{2}
\end{array}
\right ).
\eqno{(12)}
$$
After symmetry breaking, only $U(1)_Q$ and $U(1)_M$ symmetries are
preserved.
\\

In eq(8), parameters $f^{(j)}$ and $F=(f^{(jk)})$
are selected as:
$$
\begin{array}{l}
F = \frac{\sqrt{2}}{\mu} K
\left (
\begin{array}{ccc}
M_{\nu e} &&  \\
& M_{\nu \mu} &  \\
&& M_{\nu \tau}
\end{array}
\right )
K^{\dag}  \\
f^{(1)} = \frac{\sqrt{2} M_{e}}{ \mu},
f^{(2)} = \frac{\sqrt{2} M_{\mu}}{ \mu},
f^{(3)} = \frac{\sqrt{2} M_{\tau}}{ \mu}.
\end{array}
\eqno{(13)}
$$
\\

Gauge fields $F_{1 \mu}, F_{2 \mu}, B_{1 \mu}$,
$B_{2 \mu}$ and $B_{3 \mu}$ are not
eigenvectors of mass matrix. In order to obtain the eigenvectors of
mass matrix, three sets of field transformations are needed:
$$
\begin{array}{l}
B_{4 \mu}= {\rm cos}\beta B_{1 \mu} - {\rm sin}\beta B_{2 \mu}  \\
B_{5 \mu}= {\rm sin}\beta B_{1 \mu} + {\rm cos}\beta B_{2 \mu} ,
\end{array}
\eqno{(14)}
$$
$$
\begin{array}{l}
W_{\mu}={\rm cos}\alpha F_{1 \mu}+{\rm sin}\alpha F_{2 \mu}  \\
W_{2 \mu}=-{\rm sin}\alpha F_{1 \mu}+{\rm cos}\alpha F_{2 \mu}  \\
C_{1 \mu}={\rm cos}\alpha B_{4 \mu}+{\rm sin}\alpha B_{3 \mu}  \\
C_{2 \mu}=-{\rm sin}\alpha B_{4 \mu}+{\rm cos}\alpha B_{3 \mu} ,
\end{array}
\eqno{(15)}
$$
and
$$
\begin{array}{l}
Z_{\mu}= {\rm sin}\theta _W C_{1 \mu}-{\rm cos}\theta _W W^3_{ \mu}   \\
A_{\mu}= {\rm cos}\theta _W C_{1 \mu}+{\rm sin}\theta _W W^3_{ \mu}    \\
Z_{2 \mu}= {\rm sin}\theta _W C_{2 \mu}-{\rm cos}\theta _W W^3_{2 \mu}  \\
A_{2 \mu}= {\rm cos}\theta _W C_{2 \mu}+{\rm sin}\theta _W W^3_{2 \mu}.
\end{array}
\eqno{(16)}
$$Because these transformations are pure algebra operations, they do not
affect
the symmetry of the Lagrangian.
\\

Finally, the Lagrangian density changed into the following form:
$$
\begin{array}{l}
{\cal L}_l +{\cal L}_{v-l}=
 - \sum_{j=1}^3 \overline{e}^{(j)} (\gamma ^{\mu}
\partial _{\mu}+ M^{(j)} ) e ^{(j)} \\
- \sum_{j=1}^3 \overline{\nu}^{(j)}
( \gamma ^{\mu} \partial _{\mu} + M^{(j)}_{\nu} ) \nu^{(j)}  \\

+\frac{1}{2} \sqrt{g^2 + g'^2} {\rm sin}2\theta_W
j^{em}_{\mu}
 ( {\rm cos}\alpha A^{\mu}- {\rm sin}\alpha  A^{\mu}_2  )  \\

- \sqrt{g^2 + g'^2} j^{z}_{\mu}
( {\rm cos}\alpha Z^{\mu} - {\rm sin}\alpha Z_2^{ \mu} ) \\

+ \frac{\sqrt{2}}{2} ig
(\sum_{j=1}^3 \overline{\nu}^{(j)}_{\theta L} \gamma ^{\mu} e^{(j)}_L)
( {\rm cos}\alpha W_{\mu}^{+} - {\rm sin}\alpha W_{2 \mu}^{+} )  \\

+ \frac{\sqrt{2}}{2} ig
(\sum_{j=1}^3 \overline{e}^{(j)}_L \gamma ^{\mu} {\nu}^{(j)}_{\theta L} )
( {\rm cos}\alpha W_{\mu}^{-} - {\rm sin}\alpha W_{2 \mu}^{-} )  \\

- \frac{i}{2} g' {\rm sin}\beta  \sum_{j=1}^3
( \overline{e}^{(j)}_R \gamma ^{\mu} e ^{(j)}_R
- \overline{\nu}^{(j)}_R \gamma^{\mu} \nu^{(j)}_R  ) \\
\cdot (- {\rm sin}\beta {\rm cos}\alpha ({\rm sin}\theta_W Z_{\mu}
 +{\rm cos}\theta_W A_{\mu} ) \\
+{\rm sin}\beta {\rm sin}\alpha
({\rm sin}\theta_W Z_{2 \mu} +{\rm cos}\theta_W A_{2 \mu} )+
{\rm cos}\beta B_{5 \mu} )
\end{array}
\eqno{(17)}
$$
$$
\begin{array}{l}
{\cal L}_g = -\frac{1}{2}  W^{+ \mu \nu}_0  W^{-}_{0 \mu \nu}
-\frac{1}{4}  Z^{\mu \nu} Z_{ \mu \nu}
-\frac{1}{4}  A^{\mu \nu} A_{ \mu \nu}
 \\
-\frac{1}{2}  W^{+ \mu \nu}_{2 0}  W^{-}_{2 0 \mu \nu}
-\frac{1}{4}  Z^{\mu \nu}_2  Z_{2  \mu \nu}
-\frac{1}{4}  A^{\mu \nu}_2  A_{2  \mu \nu}
\\
-\frac{1}{4}  B^{\mu \nu}_5  B_{5  \mu \nu}
 -\frac{\mu ^2}{2}  Z^{\mu }  Z_{ \mu }
-\mu ^2 {\rm cos}^2 \theta _W  W^{+ \mu}  W^{-}_{\mu} +{\cal L}_{g I}
\end{array}
\eqno{(18)}
$$
where, ${\cal L}_{g I}$ only contains interaction terms of gauge fields.
In the above relations, all strengths of gauge
fields have the following form:
$$
A_{\mu \nu} = \partial _{\mu} A_{\nu} - \partial _{\nu} A_{\mu}
\eqno{(19)}
$$
where $A$ represents $W,~W_2,~Z,~Z_2,~A,~A_2$, or $B_5$.
The definitions of some other arguments are:
$$
\begin{array}{l}
W^{\pm}_{m \mu } = \frac{1}{\sqrt{2}} (W^1_{m \mu} \mp i W^2_{m \mu} )
\\
~~~~~~~~~~( m=1,2, ~W_{1 \mu} \equiv W_{\mu}  ) ,  \\
j_{ \mu }^{em} = -i (\overline{e} \gamma _{\mu} e
+\overline{\mu} \gamma _{\mu} \mu +\overline{\tau} \gamma _{\mu} \tau) ,\\
j_{ \mu }^{Z} = j^3_{\mu} - {\rm sin}^2 \theta_W  j^{em}_{\mu}.
\end{array}
\eqno{(20)}
$$
\\

From the above results, it could be seen that, all leptons are massive,
gauge bosons $W^{\pm}$ and $Z$ are also massive, but gauge fields
$A, A_2, Z_2, W_2^{\pm}$ and $B_5$ are massless. The mass relation between
$W^{\pm}$ and $Z$ is similar to that in the Standard Model.
The current structures in this
model, especially the charged current structures,
 are also similar to  those in the standard model.
Besides, it could be noticed that, there exist two different
electromagnetic fields $A_{\mu}$ and $ A_{2 \mu}$, so there exist
two different coupling constants of electromagnetic interactions.
$$
e_1 = \frac{g  g'}{\sqrt{g^2 + {g'} ^2}} {\rm cos}\alpha
~,~~
e_2 = \frac{g  g'}{\sqrt{g^2 + {g' }^2}} {\rm sin}\alpha .
\eqno{(21)}
$$
The real electromagnetic field in Nature is the mixture of
these two electromagnetic fields $A_{\mu}$ and $A_{2 \mu}$,
and the effective coupling
constant of the electromagnetic interactions should be
$$
e^2 = e_1^2 +e_2^2
~,~~
e= \frac{g  g'}{\sqrt{g^2 + {g' }^2}} .
\eqno{(22)}
$$
So the value of the parameter
$\alpha$ doesn't  affect the value of the
effective coupling constant of  electromagnetic interactions.   \\

It is expected that the parameter $\alpha$ and $\beta$ should be small,
so the contributions come from massless gauge fields
and right-hand neutrinos will be very small and the
present model will not contradict with high energy experiments. \\

It is generally believed that neutrinos detected by experiments are
left-hand neutrinos and right-hand anti-neutrinos. In this model, though
right-hand neutrinos are introduced, but they only couple with neutral
currents and the corresponding coupling constant is about
${\rm sin}^2 \beta \times {\rm ordinary~weak~coupling~constant}$.
Because ${\rm sin} \beta$ is very small, any
process contains right-hand neutrinos will be extremely weak.
Therefore, neutrinos detected by
experiments are mostly left-hand neutrinos, and right-hand neutrinos is very
hard to be detected. The Nature still has no left-right symmetry and
the P Parity is not conserved. Besides, right-hand neutrinos have no direct
coupling with charged leptons,
so  almost all neutrinos generated from $\beta$ decay are
left-hand neutrinos.
That make them more difficult to be found in experiments.
\\

Because of neutrinos' mixing, the lepton-number is not separately conserved.
But no total lepton-number violating term is introduced, so the total
lepton-number is still conserved. In another word,
one type of lepton can convert
into another type of lepton through neutrino mixing,
but the total lepton number is not changed in this process.
So, any process which will change the total
lepton number, such as double beta decay, is prohibited by
this model.  This selection rule originate from $U(1)_M$ symmetry.  \\

Though lepton-number is not separately conserved, some
lepton-number-violating processes, such as
$\mu^- \to e^- \gamma$, $\tau^- \to \mu^- \gamma$ and
$\tau^- \to e^- \gamma$, will be extremely weak.
As an example, let's discuss $\mu^- \to e^- \gamma$.
In this process, all neutrinos appear as inner lines of a Feynman diagram.
Suppose that this $\mu^-$ changes into a left-hand neutrino
and a $W^-$ particle and both of them are
inner lines of a Feynman diagram.
The Feynman propagater for neutrino is
$<0| \overline{\nu} \nu |0> =
<0| \overline{\nu}_L \nu_R |0> + <0| \overline{\nu}_R \nu_L |0>$.
If one end of propagater is $\overline{\nu}_L$, another end must be
$\nu_R$. But right-hand neutrinos have no direct interactions with
electrons,
nor with $W^-$ particle, it has to interact with
a neutral gauge boson to become $\overline{\nu}_R$ which is at one
end of a new neutrino inner line. The other end of
this neutrino inner line must be $\nu_L$ which interact with $W^-$ to
produce an electron. This diagram is at least at one-loop level.
Besides, the right-hand neutrinos have very weak coupling with gauge
bosons. Both factors will make the branch ratio
of this process extremely small. \\

In this model, two things are accomplished  simultaneously: the
introduction of neutrino mass and the disappearing of Higgs particle through
introducing a new but familiar charge --- total lepton number and total
baryon number, 
 and another set of gauge fields.
The structure of this model is completely determined by the
symmetry of the model.
This model is renormalizable[7].
If we did not add the restriction of minimal on
the theory, we could also construct a lot of models which contain massive
neutrinos and even the Higgs particle[7].
A minimum model that contains both
massive neutrinos and Higgs particle is that[7]:
$$
{\cal L} = {\cal L} _l + {\cal L} _g + {\cal L} _{H-l} + {\cal L} _H
\eqno{(23)}
$$
where ${\cal L} _H$ is the Lagrangian for Higgs fields which is the same as
that in the Standard Model,
${\cal L} _l$ is given by eq(6),
${\cal L}_{H-l}$ is given by eq(8) but should replace vacuum potential with
Higgs fields, and ${\cal L} _g$ is given by:
$$
{\cal L}_g = -\frac{1}{4}  F^{i \mu \nu}_1 F^i_{1 \mu \nu}
-\frac{1}{4}  B^{\mu \nu}_1 B_{1 \mu \nu}
-\frac{1}{4}  B^{\mu \nu}_2 B_{2 \mu \nu}.
\eqno{(24)}
$$
This model has some interesting properties but has more independent
parameters than the previous one[7]. Besides, its structure is relatively
more loosy.
\\

\section*{Reference:}
\begin{description}
\item[\lbrack 1 \rbrack] Fukuda Y, et al. Phys. Rev. Lett. 81: 1562-1567
(1998)
\item[\lbrack 2 \rbrack] Fukuda Y, et al. Phys. Rev. Lett. 81: 1158-1162
(1998)
\item[\lbrack 3 \rbrack] Fukuda Y, et al. Phys. Lett. B436: 33-41 (1998)
\item[\lbrack 4 \rbrack] Fukuda Y, et al. Phys. Lett. B433: 9-18(1998)
\item[\lbrack 5 \rbrack] Wolfenstein L. Phys. Rev. D17: 2369 (1978), ~~
Phys. Rev. D20: 2634 (1979)
\item[\lbrack 6 \rbrack] Mikheyev SP, Smirmov AYu. Sov. J. Nucl. Phys.
42:913~(1986),~~ Sov. Phys. JETP64:4~(1986),~~
Nuovo Cimento 9C:17~(1986)
\item[\lbrack 7 \rbrack] Ning Wu, Electroweak models with massive neutrinos,
in this paper, both models given in this paper are discussed in details;
Ning Wu, Gauge field theory with massive gauge fields, 
hep-ph/9805453, hep-ph/9807416
\item[\lbrack 8 \rbrack]  S.Glashow, Nucl Phys {\bf 22}(1961) 579
\item[\lbrack 9 \rbrack]  S.Weinberg, Phys Rev Lett {\bf 19} (1967) 1264
\item[\lbrack 10 \rbrack]  A.Salam, in Elementary Particle Theory,
eds. N.Svartholm (Almquist and Forlag, Stockholm,1968)
\end{description}

\end{document}